\documentclass[conference]{IEEEtran}

\usepackage{bbm}
\usepackage{amsmath}
\usepackage{acronym}  
\usepackage[dvips]{color}
\usepackage{epsf}
\usepackage{times}
\usepackage{epsfig}
\usepackage{notoccite}
\usepackage{graphicx}
\usepackage{epstopdf}
\usepackage{pstricks}
\usepackage{amssymb}
\usepackage{amsxtra}
\usepackage{here}
\usepackage{rawfonts}
\usepackage{float}
\usepackage{times}
\usepackage{url}
\usepackage{cite}
\usepackage{caption}
\usepackage{subcaption}
\usepackage{algorithm}
\usepackage{algpseudocode}
\usepackage{blindtext}
\usepackage{enumitem}
\usepackage{xcolor,cite,etoolbox}
\usepackage{relsize}
\usepackage{lipsum}
\usepackage{graphicx}
\usepackage{tabularx}
\usepackage{xparse}
\usepackage{array}
\newcolumntype{P}[1]{>{\centering\arraybackslash}p{#1}}
\usepackage{mleftright}

\usepackage{mathtools}

\usepackage{graphics}
\usepackage{physics}
\usepackage{amssymb}
\usepackage{siunitx}
\include{newcommands}
\usepackage{multicol}

\usepackage[nomain,acronym,shortcuts]{glossaries}
\makeglossaries
\newcommand*{\acro}[3][]{\newacronym[#1]{#2}{#2}{#3}}

\acro{D2D}{device-to-device}
\acro{SIR}{signal-to-interference-ratio}
\acro{SINR}{signal-to-interference-plus-noise-ratio}
\acro{PCP}{Poisson cluster process}
\acro{CoMP}{coordinated multi-point}
\acro{BS}{base station} 
\acro{MD-CoMP}{macrodiversity CoMP transmission}
\acro{MAC}{medium-access-control}
\acro{JT-CoMP}{joint transmission CoMP}
\acro{CoMP-JT}{coordinated multipoint joint transmission}
\acro{SBS}{small base station}
\acro{MDSD}{multiple devices to a single device}
\acro{MDS}{maximum distance separable}
\acro{SCN}{small cell network}
\acro{PPP}{Poisson point process}
\acro{TCP}{Thomas cluster process}
\acro{CSI}{channel state information}
\acro{PDF}{probability distribution function}
\acro{PMF}{probability mass function}
\acro{RV}{random variable}
\acro{i.i.d.}{independently and identically distributed}
\acro{MBMS}{multimedia broadcasting multicasting service}
\acro{EE}{energy efficiency}
\acro{HCP}{hard-core placement}
\acro{CCDF}{complementary cumulative distribution function}
\acro{CDF}{cumulative distribution function}
\acro{PC}{probabilistic caching}
\acro{RC}{random caching}
\acro{CPF}{caching popular files} 
\acro{PGFL}{probability generating functional}
\acro{KKT}{Karush-Kuhn-Tucker}
\acro{PGF}{point generating function}
\acro{SCA}{successive convex approximation}
\acro{HD}{high-definition}
\acro{FHD}{full-high-definition}
\acro{UHD}{ultra-high-definition}
\acro{VR}{virtual reality}
\acro{AR}{augmented reality}
\acro{5G}{fifth-generation}
\acro{QoS}{quality-of-service}
\acro{QoE}{quality-of-experience}
\acro{IoT}{internet of things}
\acro{MHCPP}{Matern hardcore point process}
\acro{LoS}{line-of-sight}
\acro{NLoS}{non-line-of-sight}
\acro{PSD}{power spectral density}
\acro{MEC}{mobile edge computing}
\acro{E2E}{end-to-end}
\acro{THz}{terahertz}
\acro{CLT}{central limit theorem}
\acro{HQ}{High Quality}
\acro{eMBB}{enhanced mobile broadband}
\acro{URLLC}{ultra reliable low latency communications}
\acro{mmWave}{millimeter wave}
\acro{EVT}{extreme value theory}
\acro{GEV}{generalized extreme value}
\acro{LIS}{large intelligent surface}
\acro{RIS}{reconfigurable intelligent surface}
\acro{RF}{radio frequency}
\acro{UE}{user equipment}
\acro{MIMO}{multiple-input multiple-output}
\acro{EVaR}{entropic value-at-risk}
\acro{DNN}{deep neural network}
\acro{MDP}{Markov decision process}
\acro{RL}{reinforcement learning}
\acro{RNN}{recurrent neural network}
\acro{ANN}{artificial neural networks}
\acro{LSTM}{long short-term memory}
\acro{ReLu}{rectified linear unit}
\acro{VaR}{value-at-risk}
\acro{SNR}{signal-to-noise ratio}
\acro{AoSA}{array of subarray}
\acro{XR}{extended reality}
\acro{AoA}{angle of arrival}
\acro{ULA}{uniform linear array}
\acro{AoD}{angle of departure}
\acro{EM}{electromagnetic}
\acro{HRLLC}{s high-rate and high-reliability low latency communications}
\acro{6DoF}{six degrees of freedom}
\acro{MR}{mixed reality}
\acro{PAPR}{peak to average power ratio}
\acro{OFDM}{orthogonal frequency-division multiplexing}
\acro{OFDMA}{orthogonal frequency-division multiple access}
\acro{SC-FDM}{single carrier frequency-division multiplexing}
\acro{ToA}{time of arrival}
\acro{MUSIC}{multiple signal classification}
\acro{IoE}{Internet of Everything}

\usepackage{datetime}
\usepackage{amssymb}
\usepackage{subcaption}
\usepackage{verbatim}

\newtheorem{proposition}{\bf Proposition}

\begin{document}
	\title{Joint Sensing and Communication for Situational Awareness in Wireless THz Systems \vspace{-.5cm}} \vspace{-1cm}
	\author{\IEEEauthorblockN{\small{Christina Chaccour\IEEEauthorrefmark{1},
		Walid Saad \IEEEauthorrefmark{1},
		Omid Semiari \IEEEauthorrefmark{2},
		Mehdi Bennis\IEEEauthorrefmark{3}, 
		and Petar Popovski \IEEEauthorrefmark{4}}}
	\IEEEauthorblockA{\small{\IEEEauthorrefmark{1}Wireless@ VT, Bradley Department of Electrical and Computer Engineering, Virginia Tech, Arlington, VA, USA,}}
	\IEEEauthorblockA{\small{\IEEEauthorrefmark{2}Department of Electrical and Computer Engineering, University of Colorado, Colorado Springs, CO, USA,}}
	\IEEEauthorblockA{\small{\IEEEauthorrefmark{3}Centre for Wireless Communications, University of Oulu, Finland,}}
	\IEEEauthorblockA{\small{\IEEEauthorrefmark{4} Department of Electronic Systems, Aalborg University, Denmark.}}
	\IEEEauthorblockA{\small{Emails:\{christinac, walids\}@vt.edu, osemiari@uccs.edu, mehdi.bennis@oulu.fi, petarp@es.aau.dk }}\vspace{-.9cm}}
	\maketitle
\normalsize
	\begin{abstract}
		Next-generation wireless systems are rapidly evolving from communication-only systems to multi-modal systems with integrated sensing and communications. In this paper a novel joint sensing and communication framework is proposed for enabling wireless \ac{XR} at \ac{THz} bands. To gather rich sensing information and a higher \ac{LoS} availability, \ac{THz}-operated \acp{RIS} acting as base stations are deployed. The sensing parameters are extracted by leveraging \ac{THz}'s quasi-opticality and opportunistically utilizing uplink communication waveforms. This enables the use of the same waveform, spectrum, and hardware for both sensing and communication purposes. The environmental sensing parameters are then derived by exploiting the sparsity of \ac{THz} channels via tensor decomposition. Hence, a high-resolution indoor mapping is derived so as to characterize the spatial availability of communications and the mobility of users. Simulation results show that in the proposed framework, the resolution and data rate of the overall system are positively correlated, thus allowing a joint optimization between these metrics with no tradeoffs. Results also show that the proposed framework improves the system reliability in static and mobile systems. In particular, the highest reliability gains of $\SI{10}{\%}$ are achieved in a walking speed mobile environment compared to communication only systems with beam tracking.
	\end{abstract}
	\begin{IEEEkeywords}
		extended reality (XR), terahertz (THz),  reliability, sensing, joint sensing and communications.
	\end{IEEEkeywords}
	\IEEEpeerreviewmaketitle
	\vspace{-.3cm}
\section{Introduction}\label{sec:Intro}
\vspace{-.2cm}
The sixth generation (6G) of  wireless systems is expected to support radically  intelligent and autonomous services in an \ac{IoE} system \cite{saad2019vision}. Supporting such applications requires versatile wireless capabilities that go beyond communications to encompass sensing, localization, and control. Such versatile capabilities can enable many applications such as \ac{XR}\footnote{\ac{XR} encompasses \ac{AR}, \ac{MR}, and \ac{VR}.}. In fact, successfully designing a versatile wireless XR system that delivers a multi-sensory immersive experience faces several networking challenges. First, to transmit high-definition $360^\circ$ \ac{XR} content, the network must guarantee extremely high data rates beyond what is supported by modern-day systems\footnote{Early generations of \ac{XR} services can be delivered using mmWave systems. Nonetheless, 5G remains limited by a maximum downlink data rate up to a few $\SI{}{Gbps}$. Meanwhile, to guarantee the requirements of the new generation of \ac{XR} services, higher data rates are needed, thus, this naturally motivates the migration towards the \ac{THz} band ($0.1-\SI{10}{THz}$).}. Second, along with these data rates, providing reliable (multi-sensory) haptic \ac{XR} communications also requires maintaining near-zero \ac{E2E} latency, which cannot be achieved even at the 5G \acrfull{mmWave} bands. Third, to support wireless \ac{XR}, there is a need for instantaneous tracking of the \ac{6DoF} of the head and body of each user along with a cognizant situational awareness of the XR users' surroundings. Empowering a wireless network with these capabilities can be done by leveraging higher frequency bands, namely the terahertz (THz) band ($0.1-\SI{10}{THz}$). Owing to their abundant bandwidth and directional transmission, THz bands can deliver extremely high data rates (order of Tbps), and, simultaneously, they can provide a high-resolution environmental sensing capability (at the centimeter level) \cite{sarieddeen2021overview}.\\
\indent This joint sensing and communication capability of \ac{THz} systems opens the door for a mutual feedback between the communication and sensing functions. Particularly, the sensing input can be exploited for two purposes: 1) Sensing information is needed to enhance the service intelligence (used to precisely teleport users in collaborative XR, e.g., a multiplayer \ac{XR} game setup), and, most importantly, 2) Sensing a dynamic environment allows wireless networks to be \emph{cognizant of the environment's stochasticity in real-time, thus, achieving operational intelligence}. For instance, \ac{THz} beams are narrow, and they are highly susceptible to changes induced by the mobility of users. To overcome this challenge, beam training and channel estimation of mobile \ac{UE} should be performed more frequently. However, this causes significant overhead and delays, and it falls short of delivering stringent \ac{IoE} requirements. Meanwhile, a joint sensing and communication system that shares the waveform, hardware, and spectrum can help overcome the aforementioned challenges by achieving a higher spectral efficiency.\\
\indent Nonetheless, designing joint sensing and communication systems faces multiple challenges, as sensing and communication are two operations that are functionally different.\footnote{For instance, sensing typically relies on unmodulated signals or short pulses and chirps. Communication signals, on the other hand, are a mix of unmodulated (pilots) and modulated signals.} Hence, to successfully leverage high-rate communication links for sensing purposes, simultaneously and in real-time, we must answer the following fundamental questions:
\begin{enumerate}[label=(\alph*)]
	\item Given a particular resolution, how much environmental sensing information can we extract from \ac{THz} communication links, and how?
	\item Can we perform such an operation in near real-time without adding a substantial latency to the network's \ac{E2E} delay? 
	\item What are the gains achieved with such information in terms of minimizing frequent beam training and channel estimations? (to fulfill wireless \ac{XR} requirements)
\end{enumerate} 
\vspace{-0.25cm}
\subsection{Prior Works}
The concept of joint communication and sensing has recently seen a surge in the literature \cite{ahmadipour2021information, zhang2021design, buzzi2019using, elbir2021terahertz, guerra2021real}. In \cite{ahmadipour2021information}, the authors studied the problem of joint sensing and communication in memoryless state-dependent channels. The work in \cite{zhang2021design} designed a joint sensing and communication integrated system to support these dual functions for 5G \ac{mmWave} bands. The authors in \cite{buzzi2019using} analyzed the performance of a massive \ac{MIMO}-radar \acrfull{BS} jointly using the same spectrum for communication and sensing. However, the work in \cite{ahmadipour2021information} relies on an information-theoretic construct whose analysis is limited to a single transmitter and receiver. Here, the solutions of \cite{zhang2021design} and \cite{buzzi2019using} cannot perform sensing and communication within the same spatial-temporal frames given that they multiplex sensing and communication in time or space. Furthermore, the works in \cite{ahmadipour2021information, zhang2021design, buzzi2019using} do not leverage the \ac{THz} band's quasi-opticality. Meanwhile, the authors in \cite{elbir2021terahertz} and \cite{guerra2021real} investigated the sensing capabilities of \ac{THz} bands. For instance, in \cite{elbir2021terahertz} model-based and model-free hybrid beamforming techniques were investigated for a joint radar and communication system. The work in  \cite{guerra2021real} considered the join detection, mapping and navigation problem by a drone with real-time learning capabilities. Nonetheless, these works \cite{ahmadipour2021information, zhang2021design, buzzi2019using, elbir2021terahertz, guerra2021real} do not opportunistically utilize the same spectrum, waveform and hardware at \ac{THz} for joint sensing and communications Consequently, such joint deployment strategies lead to higher costs in terms of hardware and resources, which is not suitable for commercial \ac{XR} services. In fact, the works in \cite{ahmadipour2021information, zhang2021design, buzzi2019using, elbir2021terahertz, guerra2021real} do not study  the extraction of sensing parameters from uplink communication signals. Such sensing parameters, if properly extracted and estimated at the \ac{THz} bands, can provide new opportunities for system performance enhancements.\\
\indent The main contribution of this paper is, thus, a novel joint sensing and communication framework for \ac{XR} systems leveraging \ac{THz}-operated \acrfullpl{RIS}. In our considered network, the \ac{THz} uplink communication waveform is used to both deliver high data rate \ac{XR} content and opportunistically extract environmental sensing parameters. These parameters allow us to reduce the constant need for beam training in the highly varying \ac{THz} channel. Also, this framework enables performing high-resolution sensing while harnessing a communication system's resources, waveform, spectrum, and hardware. In particular, we propose a novel approach that uses tensor decomposition to leverage the sparsity of the \ac{THz} channel and guarantees extracting \emph{a unique solution} for the environmental sensing parameters. This approach allows near real-time processing of an indoor high-resolution situational-awareness map that is then used to localize active users, asses the blockage score, and consequently identify the spatial availability of \acrfull{LoS} communication links between \ac{XR} users and their \acp{RIS} unit. To our best knowledge, \emph{this is the first work that performs interleaved user-environment tracking while harnessing the uplink signal at \ac{THz} frequency bands for wireless \ac{XR} systems}. Simulation results show that the resolution and data rate of our proposed approach are positively correlated, thus allowing us to jointly optimize both with no tradeoffs. Our proposed framework achieves a high spectral efficiency, that is, respectively, $\SI{42}{\%}$ and $\SI{75}{\%}$ superior to communication only systems with beam tracking and a joint sensing and communication system in hardware only. \\
\vspace{-0.8cm}
\section{System Model}\label{Sec:Sys-Model}
Consider the uplink of an \ac{RIS}-based single cell operating as a joint sensing and communication system in a confined indoor area. A set $\mathcal{B}$ of $B$ \acp{RIS} acting as \ac{THz} operated \acp{BS} are transmitting \ac{XR} content and providing situational awareness for a set $\mathcal{U}$ of $U$ mobile wireless \ac{XR} users. Here, \emph{situational awareness} is a mapping of the physical world  in  terms  of  location,  orientation  and  state  of  physical objects  in  the  digital  realm  at  a  certain  level  of  correctness. In our model, on the one hand, we use \acp{RIS} to provide nearly continuous \ac{LoS} data links to \ac{XR} users. On the other hand, this \ac{RIS}-enhanced architecture enables creation of multiple independent paths capable of gathering rich information about the environment \cite{chaccour2021seven} and \cite{basar2019wireless}.
\begin{figure}[!t]
\includegraphics[width=0.63\textwidth]{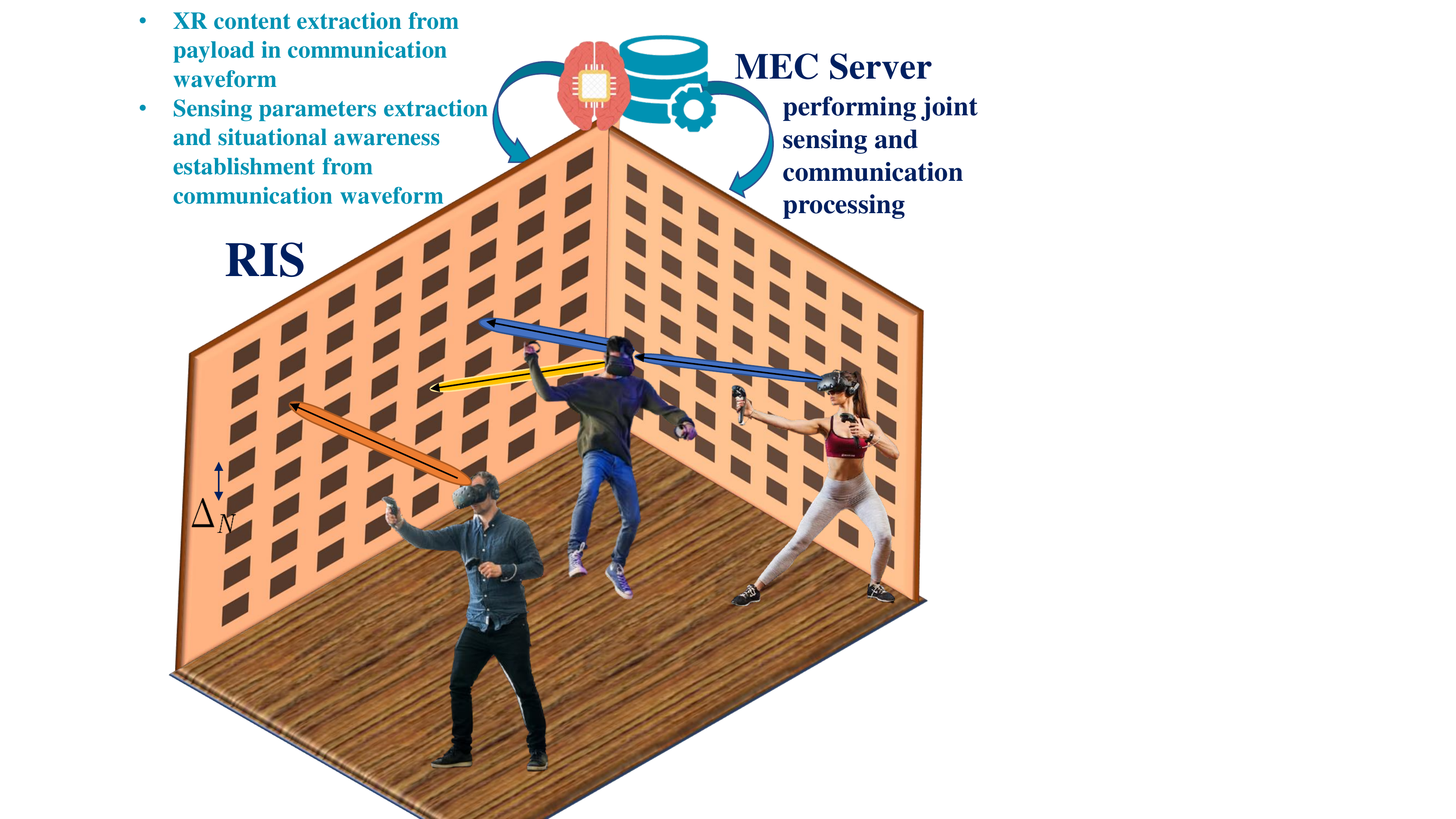}
\caption{\small{Illustrative example of the proposed THz system model for joint communication and sensing in XR applications.}}
\label{fig:model}
\vspace{-0.5cm}
\end{figure} 
Each \ac{RIS} consists of a two-dimensional large antenna array that has an \ac{AoSA} structure as in \cite{faisal2020ultramassive}. In this structure, the antenna elements of the same subarray are connected to the same phase shifter and \ac{RF} chain, thus reducing the amount of phase shifters needed compared to a fully-connected array structure. Each subarray constitutes an \emph{\ac{RIS} unit} that has a square shape and a large number of antennas $M$. The antenna spacing within each \ac{RIS} subarray is $\delta_m$. The total number of subarrays is $N$ with the spacing between subarrays being $\Delta_N$. These subarrays collectively adopt a hybrid beamforming architecture.  Without loss of generality, we assume $N \geq U$, thus providing \ac{XR} users with sufficient degrees of freedom in terms of association. This subarray structure allows us to have $N$ independent channels for a single \ac{RIS}.  The \acp{RIS} are static and their locations are known a priori by other \acp{RIS}. Let $\boldsymbol{v}^{b,n}=\left[v^{b,n}_x, v^{b,n}_y\right]^T \in \mathbb{R}^2$ be the position of the $n$th subarray of \ac{RIS} $b \in \mathcal{B}$. The \ac{XR} users are mobile and may change their locations and orientations at any point in time. Let $\boldsymbol{p}^u=\left[p^u_x, p^u_y\right] ^T \in \mathbb{R}^2$ be the position of the user $u$.
\subsection{Joint Communication and Sensing Model}
Consider an arbitrary \ac{RIS} in $\mathcal{B}$ that captures the situational awareness and mapping information, and receives high data rate \ac{XR} content as shown in Fig.~\ref{fig:model}. We assume that the \ac{RIS} has performed synchronization and initial beam alignment according to existing techniques (e.g. \cite{boulogeorgos2019performance}). To maintain high-resolution real-time user tracking of the users and a situational awareness of the indoor area, the \ac{RIS} will continuously collect a sequence of $T$ snapshots of the received uplink communication signal. These snapshots enable the \acp{RIS} to estimate and track the \ac{AoA}, \ac{AoD}, and \ac{ToA} of users as well as the location of obstacles leading to \ac{NLoS} communications. Since we use indoor \acp{RIS}, then blockages mostly occur due to bodies of the users.\\
\indent Each \ac{XR} user has a \ac{ULA} of $Q$ antennas and $R$ \ac{RF} chains. Hence, if all \ac{XR} users are active, $U$ data signaling beams are sent simultaneously to the \ac{RIS}. To mitigate the high \ac{PAPR} effect of \ac{OFDM} schemes, a \ac{SC-FDM} is used to maintain high energy efficiency at the \ac{UE}. The far-field \ac{EM} wave condition is assumed to be met. After beamforming, the transmitted signal by the \ac{UE} at subcarrier $k$ is $\boldsymbol{s}_{k}(t)=\boldsymbol{F}_{\mathrm{R}}(t) \boldsymbol{F}_{k}(t) \boldsymbol{x}_{k}(t), \quad \forall k=1, \ldots, K$, where $\boldsymbol{x}_{k}(t) \in \mathbb{C}^{l\times1}$ is the transmitted uplink communication signal, $\boldsymbol{F}_{k}\in \mathbb{C}^{R\times l}$ is the digital precoding matrix for subcarrier $k$, and $\boldsymbol{F}_{\mathrm{R}}(t)\in \mathbb{C}^{Q\times R}$ is the \ac{RF} precoder for all subcarriers. After processing the communication waveform of subcarrier $k$, the uplink received signal at each \ac{RIS} subarray $n$ will be:\vspace{-.45cm}
\begin{equation}
y_{n,k}(t)=\boldsymbol{w}_{n,k}^{T} \boldsymbol{H}_{n,u}^k(t) \boldsymbol{s}_{k}(t)+\nu_{n, k}(t),
\end{equation}
where $\boldsymbol{w}_{n,k} \in \mathbb{C}^{M\times 1}$ is the combining vector of \ac{RIS} subarray $n$ at subcarrier $k$, $\boldsymbol{H}_{n,u}^k \in \mathbb{C}^{M \times Q}$ is the uplink channel matrix for subcarrier $k$, and $\nu_{n,k} \sim \mathcal{C}\mathcal{N}\left(\boldsymbol{0}, \boldsymbol{I}_{M\times Q}\right)$ is the additive Gaussian noise vector at subarray $n$ and subcarrier $k$.
To construct a single realization  for sensing the wireless environment, $T$ snapshots are recorded by the \ac{RIS}, each of which captures $J$ uplink measurements:
\begin{equation}\label{RSS}
\boldsymbol{y}_{n,k}(t)=\boldsymbol{W}_{n,k}^{T} \boldsymbol{H}_{n,u}^k(t) \boldsymbol{F}_{\mathrm{R}}^{m}(t) \boldsymbol{F}_{k}(t) \boldsymbol{s_k}(t)+\boldsymbol{\nu}_{n,k}(t),
\end{equation} 
where $\boldsymbol{y}_{n,k}(t)  \triangleq\left[
y_{n,k, 1}(t)  \ldots  y_{n,k, J}(t)\right]^{T}, \boldsymbol{\nu}_{n,k}(t)  \triangleq\left[\nu_{n,k, 1}(t)  \ldots \nu_{n,k, J}(t)\right]^{T}, 
\boldsymbol{W_{n,k}}  \triangleq\left[
\boldsymbol{w}_{n,k,1}  \ldots  \boldsymbol{w}_{n,k,J}\right].$
Moreover, the $M \times Q$ \ac{MIMO} channel matrix between user $u$ and \ac{RIS} subarray $n$ in the time domain can be written as: $
\boldsymbol{H}_{n,u}(t)=\sum_{p=1}^{P} \alpha^p_{n,u} \boldsymbol{a_r}\left(\phi^p_{n,u}\right) \boldsymbol{a_t}^H\left(\theta^p_{n,u}\right)\delta(t-\tau_{n,u}), $
where $\left(.\right)^H$ is the conjugate transpose, $\alpha^l_{n,u} \in \mathbb{C}$ is the complex channel gain, $P$ is the number of distinct spatial paths,  $\theta_{n,u}^{l}$ is the \ac{AoD} of the $p$th path of antenna $u$, $\tau_{n,u}$ is the \ac{ToA} of the path, $\phi_{n,u}^{p}$ is the \ac{AoA} of the $p$th path at subarray $n$, and  $\boldsymbol{a_t}\left(\theta^{p}_{n,u}\right)=\frac{1}{\sqrt{Q}}\left[1, e^{j \frac{2\pi}{\delta_m} \sin \left(\theta^p_{n,u}\right)}, \cdots, e^{j(Q-1) \frac{2\pi}{\delta_n} \sin \left(\theta^p_{n,u}\right)}\right]^{T}$ is the array steering vector of the antenna array of the \ac{XR} \ac{UE} $\boldsymbol{a_r}\left(\phi^{pk}_{n,u}\right)=\frac{1}{\sqrt{M}}\left[1, e^{j \frac{2\pi}{\delta_m} \sin \left(\phi^p_{n,u}\right)}, \cdots, e^{j(M-1) \frac{2\pi}{\delta_n} \sin \left(\phi^p_{n,u}\right)}\right]^{T}$ is the array steering vector of the \ac{RIS} subarray that has a \ac{ULA} structure\footnote{While uniform planar arrays can be used given the 2D structure of the subarray, we adopt a \ac{ULA} structure for analytical tractability.}
Consequently, the channel matrix in the frequency domain associated with the $k$th subcarrier is given by:\vspace{-0.25cm}
\begin{equation}\label{channel}
\boldsymbol{H}^k_{n,u}(f)=\sum_{p=1}^{P} \alpha^p_{n,u} \boldsymbol{a_r}\left(\phi^p_{n,u}\right) \boldsymbol{a_t}^H\left(\theta^p_{n,u}\right)\exp\left(-\frac{j2\pi f\tau_{n,u}}{K} \right).
\end{equation}
\indent The snapshots in \eqref{RSS} can be collected over \ac{LoS} or \ac{NLoS} communication signal waveforms. In the case of a \ac{LoS} uplink transmission, the channel gain will be \cite{chaccour2020can}: $ \alpha^L_{n,u}=\frac{c}{4\pi f r_{n,u}} e^{-\frac{k(f)r_{n,u}}{2}} e^{-j2\pi f \tau^L_{n,u}}.$
Here, $c$ is the speed of light, $k(f)$ is the overall molecular absorption coefficients of the medium at \ac{THz} band, $f$ is the operating frequency, and $r_{n,u}$ is the distance between the \ac{XR} user $u$  and the \ac{RIS} subarray $n$. From this channel gain, we can see that in \ac{LoS} conditions, the parameters to be estimated are $\{\theta^L_{n,u}, \phi^L_{n,u}, \tau^L_{n,u}\}$. This process allows tracking \ac{XR} users in real-time. Here, we do not consider \ac{NLoS} links for communications purposes due to the poor propagation characteristics of \ac{THz} signals (e.g., limited reflection) and their high susceptibility to blockage. That is, such a \ac{NLoS} link cannot meet the high-rate \ac{XR} needs. Nonetheless, from a sensing standpoint, instead of discarding the \ac{NLoS} signal completely, we use it to track the user and gather a situational awareness of the blockage incidence. In \ac{NLoS} conditions, the channel gain is given by \cite{moldovan2014and}: 
\begin{align}\label{eqNLoS}
\alpha^N_{n,u}=\frac{c}{4\pi f (r^{(1)}_{n,u}+r^{(2)}_{n,u})} e^{\left({-\frac{k(f)(r^{(1)}_{n,u}+r^{(2)}_{n,})}{2}}\right)} R(f)e^{-j2\pi f \tau^N_{n,u}},
\end{align}
\normalsize
where  $r^{(1)}_{n,u}$ is the distance between the \ac{XR} user $u$ and the reflecting point,  and $r^{(2)}_{n,u}$ is the distance between the reflecting point and the \ac{RIS} subarray $n$. For reflections, we consider the transverse electric part of the \ac{EM} wave, that is, the signal is assumed to be perpendicular to the plane of incidence\footnote{Extending our model to a transverse magnetic component (whereby the signal is parallel to the plane of incidence) can be done similarly.}. This assumption is valid given the placement of our \ac{RIS} as shown in Fig.~\ref{fig:model}. As such, $R_{n,u}(f)=\gamma_{n,u}(f)\rho_{n,u}(f)$ is the reflection coefficient, where $\gamma(f)\approx-\exp \left(\frac{-2 \cos \left(\psi_{n,u}\right)}{\sqrt{\eta(f)^{2}-1}}\right)$ is the Fresnel reflection coefficient and $\rho_{n,u}(f)=\exp \left(-\frac{8 \pi^{2} f^{2} \sigma^{2}  \cos ^{2}\left(\psi_{n,u}\right)}{c^{2}}\right)$ is the Rayleigh factor that characterizes the roughness effect. $\psi_{n,u}$ is the angle of the incident signal to the reflector, $\eta(f)$ is the refractive index, and $\sigma$ is the surface height standard deviation. Thus, for \ac{NLoS}, the parameters to be estimated are: $\{\theta^N_{n,u}, \phi^N_{n,u}, \tau^N_{n,u}, \psi_{n,u}\}$. Next, we formulate the problem of estimating the sensing parameters from the $J$ collected snapshots in \eqref{RSS}. To do so, we leverage the sparsity of the \ac{THz} channel matrix and reformulate the received uplink signal expression as a tensor. \vspace{-0.15cm}
\section{Situational Awareness Estimation via Tensor Decomposition}
\subsection{Problem Formulation}  
Estimating the sensing parameters from $T$ communication snapshots recorded at the \ac{RIS} subarray faces multiple challenges. First, existing estimation techniques \cite{mendrzik2019enabling, ozkaptan2018ofdm, guerra2021near} rely on the collection of pilot signals and their goal is to estimate the channel. In contrast, our goal here is to collect uplink communication signals that are being used by \ac{XR} users during an \ac{XR} session so as to \emph{continuously track} the sensing parameters.\footnote{Communication was initiated after a successful initial access and beam alignment during a initial access phase prior to the \ac{XR} session.} Second, existing parameter estimation methods suffer from particular caveats: Off-grid methods like \ac{MUSIC} cannot jointly estimate multiple sensing parameters, leading to a difficult pairing problem \cite{gruber1997statistical}. Also, the \ac{MUSIC} algorithm cannot be readily applied to the peculiar \ac{AoSA} structure of hybrid \ac{THz} beamforming systems as the complexity of the method becomes impractical with large antenna arrays \cite{zhu2017hybrid}. Moreover, on-grid methods like compressive sensing are constrained by their grid spacing, which increases complexity if a high-resolution is desired.\\
\indent To successfully estimate the \ac{THz} sensing parameters, we concatenate $T$ snapshots of the received uplink signal:
\begin{equation}\label{FreqRSS}
\boldsymbol{Y}_{n,k}=\boldsymbol{W}_{n,k}^{T} \boldsymbol{H}_{n,u}^k(f) \boldsymbol{\Omega}+\boldsymbol{N}_{n,k},
\end{equation} 
\vspace{-0.65cm}
\begin{align*}
&\boldsymbol{Y}_{n,k}  \triangleq\left[\begin{array}{lll}
\boldsymbol{y}_{n,k}(1)  \ldots & \boldsymbol{y}_{n,k}(T)
\end{array}\right], \\
&\boldsymbol{N}_{n,k}  \triangleq\left[\begin{array}{lll}
\boldsymbol{\nu}_{n,k}(1) & \ldots & \boldsymbol{\nu}_{n,k}(T)
\end{array}\right],\\
&\boldsymbol{\Omega}  \triangleq\left[\begin{array}{lll}
\boldsymbol{F}_{\mathrm{R}}^{m}(1) \boldsymbol{F}_{k}(1) \boldsymbol{s_k}(1) & \ldots & \boldsymbol{F}_{\mathrm{R}}^{m}(T) \boldsymbol{F}_{k}(T) \boldsymbol{s_k}(T)
\end{array}\right]. 
\end{align*}
From \eqref{FreqRSS}, we observe that the received signal has three modes that represent the number of measurements, the number of snapshots, and the number of subcarriers, respectively. Thus, we can model it as a three-order tensor, namely, $\boldsymbol{\chi} \in \mathbb{C}^{J \times T \times K}$. In fact, substituting \eqref{channel} in \eqref{FreqRSS}:
\begin{align}\label{New_RSS}
\boldsymbol{Y}_{n,k}(t)&=\sum_{p=1}^{P} \Lambda_{n,u,k}^p\boldsymbol{\zeta}(\phi^p_{n,u})\boldsymbol{\xi}(\theta^p_{n,u})^H+ \boldsymbol{N}_{n,k}.
\end{align} 
\noindent In \eqref{New_RSS}, $\Lambda_{n,u,k}^p\triangleq\alpha^p_{n,u}\exp(-\frac{j2\pi f\tau_{n,u}}{K})$, $ \boldsymbol{\zeta_k}(\phi^p_{n,u})\triangleq\boldsymbol{W}_{n,k}^{T} \boldsymbol{a_r}\left(\phi^p_{n,u}\right)$, and $\boldsymbol{\xi_k}(\theta^p_{n,u})\triangleq\boldsymbol{\Omega}^H\boldsymbol{a_t}\left(\theta^p_{n,u}\right)$. Consequently, we can see that each  slice $\boldsymbol{Y}_{n,k}$ corresponding to the tensor $\boldsymbol{\chi} $ is a weighted sum of a common set of rank-one outer products. Hence, we can factorize the tensor as follows:
\begin{equation*}
\begin{aligned}
\boldsymbol{\chi} =\sum_{p=1}^{P} \boldsymbol{\zeta_k}(\phi^p_{n,u})\circ \boldsymbol{\xi_k}(\theta^p_{n,u})\circ \Lambda_{n,u}^p+\mathcal{W} =\left[ \left[ \mathbf{A}, \mathbf{B}, \mathbf{C} \right] \right] +\mathcal{W},
\end{aligned}
\end{equation*}
\small
\begin{equation}
\begin{aligned}
\boldsymbol{A} & \triangleq\left[\begin{array}{lll}\boldsymbol{\zeta}_1(\phi_{n,u}^P)& \ldots &\boldsymbol{\zeta}_K(\phi_{n,u}^P)\end{array}\right], \\
\boldsymbol{B} & \triangleq \left[\begin{array}{lll} \boldsymbol{\xi}_1(\theta^p_{n,u})& \ldots \boldsymbol{\xi}_K(\theta^p_{n,u})& \end{array}\right],\\
\boldsymbol{C} & \triangleq\left[\begin{array}{lll}\boldsymbol{\Lambda}_{n,u,1}^p& \ldots &\boldsymbol{\Lambda}_{n,u,K}^p \end{array}\right].
\end{aligned}
\end{equation}
\normalsize 
Here, $\circ$ is the outer product symbol, $\boldsymbol{\Lambda}_{n,u,k}^p=\left\{\alpha^p_{n,u}\exp(-\frac{j2\pi f\tau_{n,u}}{K}) \right\}_{t=1}^{T}$, and $\mathcal{W}$ is $\boldsymbol{N}_{n,k}$ in the tensor domain.
$\boldsymbol{A}, \boldsymbol{B},$ and $\boldsymbol{C}$ are the three factor matrices associated to the tensor $\boldsymbol{\chi}$. 
	\begin{figure*}[!t]
	\begin{minipage}{0.49\textwidth}
		\centering
		\includegraphics[scale=0.18]{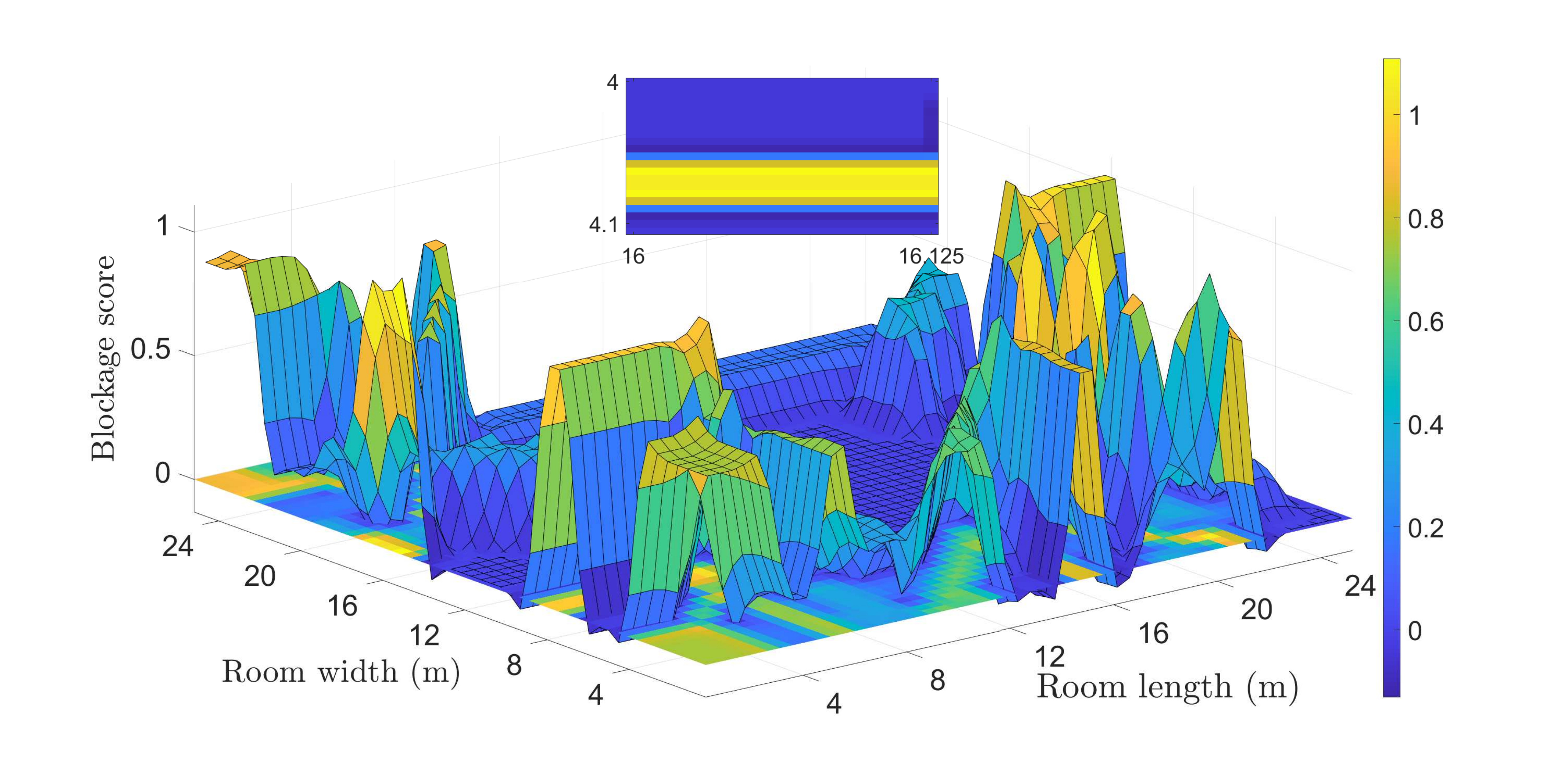}
		\subcaption{}    
	\end{minipage}
	\begin{minipage}{0.49\textwidth}
		\centering
		\includegraphics[scale=0.18]{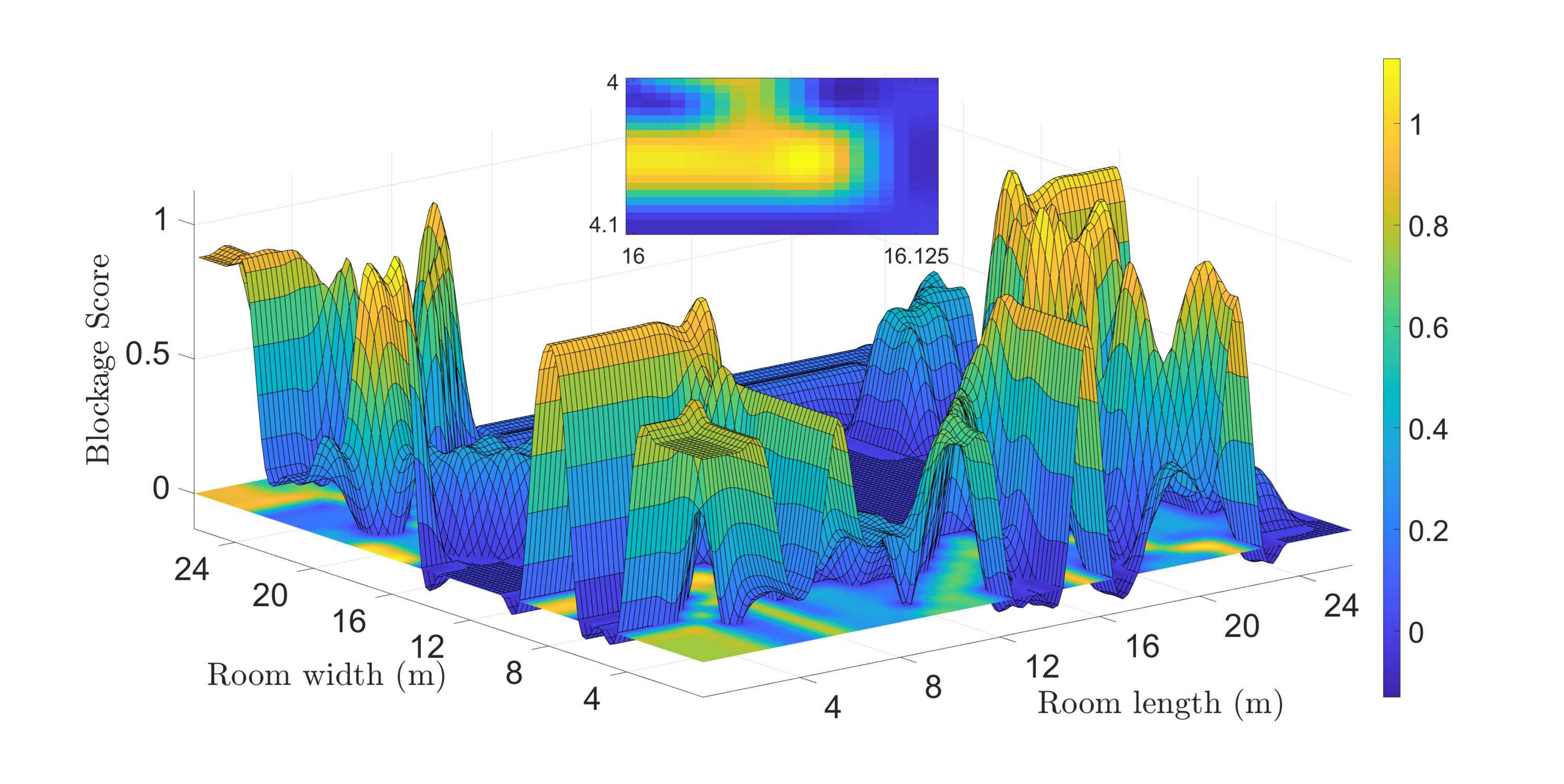}
		\subcaption{}    
	\end{minipage}
	\caption{\small{Spatial blockage score mapping at $f=\SI{0.275}{THz}$ (a) $W=\SI{7}{GHz}$, (b) $W=\SI{15}{GHz}$}, whereby a better resolution and more continuous visualizations are observed for higher bandwidths.}  
	\label{SpatialBlock}
	\vspace{-.5cm}
\end{figure*}
Here, we make some important remarks. First, owing to  the \ac{THz} channel's sparsity and the few number of paths $P$, the tensor decomposition method becomes more efficient given that the uniqueness of this decomposition is guaranteed. Second, our approach can characterize the sensing parameters from the collected snapshots without imposing additional constraints or assumptions. For instance, alternatively adopting matrix factorization almost never leads to a unique solution unless the rank of the matrix is one or further conditions are imposed on the factor matrices.
\vspace{-0.25cm}
\subsection{Proposed Sensing Parameters Estimation Method}
After factorizing the tensor $\boldsymbol{\chi}$, extracting the sensing parameters and estimating them requires solving the following optimization problem:\vspace{-0.25cm}
\small
\begin{equation}\label{optimization}
\min _{\widetilde{\boldsymbol{A}}, \widetilde{\boldsymbol{B}}, \widetilde{\boldsymbol{C}}}\left\|\boldsymbol{\chi}-\sum_{p=1}^{P} \widetilde{\boldsymbol{a}}_{p} \circ \tilde{\boldsymbol{b}}_{p} \circ \widetilde{\boldsymbol{c}}_{p}\right\|_{F}^{2},
\end{equation}
\normalsize
where $\widetilde{\mathbf{A}}=\left[\widetilde{\boldsymbol{a}_{1}} \cdots \widetilde{\boldsymbol{a}}_{P}\right], \widetilde{\boldsymbol{B}}=\left[\widetilde{\boldsymbol{b}}_{1} \cdots \widetilde{\boldsymbol{b}}_{P}\right], \widetilde{\boldsymbol{C}}=\left[\widetilde{\boldsymbol{c}}_{1} \cdots \widetilde{\boldsymbol{c}}_{P}\right]$ are the three estimated factor matrices. To solve this problem, the three factor matrices need to be estimated. To do so, we leverage the sparsity of the \ac{THz} channel that guarantees the uniqueness condition for tensor decomposition. This allows us to establish a relationship between the true factor matrices and their estimates. Thus, given these relationships we derive the environmental sensing parameters as follows:
\begin{proposition}
The \ac{AoA}, \ac{AoD}, and \ac{ToA} corresponding to path $p$ between \ac{RIS} subarray
$n$ and user $u$ are:\vspace{-.25cm}
\small
\begin{align}
\widetilde{\phi}^P_{n,u}=\arg \max _{\phi^P_{n,u}} \frac{\left|\widetilde{\boldsymbol{a}}_{k}^{H} \tilde{\boldsymbol{\zeta}}_{k}\left(\phi^P_{n,u}\right)\right|}{\left\|\widetilde{\boldsymbol{a}}_{k}\right\|_{2}\left\|\tilde{\boldsymbol{\zeta}}_{k}\left(\phi^P_{n,u}\right)\right\|_{2}}, \label{prop1}\\
\widetilde{\theta}^P_{n,u}=\arg \max _{\theta^P_{n,u}} \frac{\left|\widetilde{\boldsymbol{b}}_{k}^{H} \tilde{\boldsymbol{\xi}}_{k}\left(\theta^P_{n,u}\right)\right|}{\left\|\widetilde{\boldsymbol{b}}_{k}\right\|_{2}\left\|\tilde{\boldsymbol{\xi}}_{k}\left(\theta^P_{n,u}\right)\right\|_{2}},\label{prop2}\\
\widetilde{\tau}^P_{n,u}=\arg \min _{\tau^P_{n,u}} \frac{\left|\widetilde{\boldsymbol{c}}_{k}^{H} \tilde{\boldsymbol{\Lambda}}^P_{n,u,k}\left(\theta^P_{n,u}\right)\right|}{\left\|\widetilde{\boldsymbol{c}}_{k}\right\|_{2}\left\|\tilde{\boldsymbol{\Lambda}}^P_{n,u,k}\left(\tau^P_{n,u}\right)\right\|_{2}}.\label{prop3}
\end{align}
\normalsize
\vspace{-.25cm}
\begin{IEEEproof}
See Appendix A
\end{IEEEproof}
\end{proposition}
\eqref{prop1}-\eqref{prop3} can be solved by performing a one-dimensional search. Thus, the estimated \ac{AoA}, \ac{AoD}, and \ac{ToA} allow us to track the \ac{XR} user in near real-time (after collecting and processing the $T$ snapshots). Subsequently, after obtaining $\widetilde{\phi}^P_{n,u}, \widetilde{\theta}^P_{n,u},$ and $\widetilde{\tau}^P_{n,u}$, the attenuation factor $\alpha^P_{n,u}$ can be obtained by substitution. Moreover, the proposed Algorithm 1 enables the \ac{RIS} to track the users in real-time and characterize the spatial availability of communications. 
\begin{algorithm}[t]
	\caption{High-resolution situational awareness mapping}\label{alg:cap}
	\footnotesize
	\begin{algorithmic}
		\While{XR Session}
		\State Initialize real-time tracking mesh and blockage score mesh respectively: $\boldsymbol{E(t)}$ and $\boldsymbol{\Sigma(t)}.$
		\If {$\alpha^P_{n,u} \geq \epsilon$} 
		\State \ac{LoS} link, determine $\boldsymbol{p}^u=\left[p^u_x, p^u_y\right]^T$ from:
		\State  $\phi_{n,u}^{L} =\arctan \frac{p^u_{y}-v^{b,n}_{y}}{p^u_{x}-v^{b,n}_{x}}$ and $\tau_{n,u}^{L}=\frac{\left\|\boldsymbol{p}^u-\boldsymbol{v}^{b,n}\right\|}{c}.$
		\ElsIf{$\alpha^P_{n,u} \le \epsilon$}
			\State \ac{NLoS} link:
			\State  Determine $\psi_{n,u}$ from \eqref{eqNLoS} while using the refractive $\eta$ index
			\State of human skin. 
		\State Determine $r^{(1)}_{n,u}$ and $r^{(2)}_{n,u}$ by substituting  $\phi_{n,u}^{N}, \theta_{n,u}^{P},$ and $\psi_{n,u}$ \State into the sine law equation. 
		\State Determine obstacle location $\boldsymbol{o}^{n,u}=\left[o^{n,u}_x, o^{n,u}_y\right]^T.$ 
		\State Determine $\boldsymbol{p}^u=\left[p^u_x, p^u_y\right]$ from:
			\State  $\phi_{n,u}^{L} =\arctan \frac{p^u_{y}-o^{n,u}_{y}}{p^u_{x}-o^{n,u}_{x}}.$
			\State $r^{(1)}_{n,u}=\sqrt{\left(p^u_{x}-o^{n,u}_{x}\right)^{2}+\left(p^u_{y}-o^{n,u}_{y}\right)^{2}}.$
		\EndIf
		\For{$i=\boldsymbol{p}^u$ or $i=\boldsymbol{o}^{n,u}$}
		\State $\boldsymbol{e}_{i}(t) \leftarrow 1.$
		\State $\boldsymbol{\sigma}_i(t)  \leftarrow \boldsymbol{\sigma}_i(t-1) + 1.$
		\EndFor
		\For{$i=\boldsymbol{r}^L_{n,u}$ or $i=\boldsymbol{r}^1_{n,u}$ $i=\boldsymbol{r}^2_{n,u}$}
		\State $\boldsymbol{\sigma}_i(t)  \leftarrow \boldsymbol{\sigma}_i(t-1) + 1.$
		\EndFor
		\State Normalize blockage score map: $\bar{\boldsymbol{\Sigma}}(t)=\frac{\boldsymbol{\Sigma(t)}}{\arg \max_{\boldsymbol{\sigma}(t)}\boldsymbol{\Sigma}(t)}.$
		\EndWhile
	\end{algorithmic}
\end{algorithm}
First, owing to the high gap between \ac{LoS} and \ac{NLoS} links, the channel gain estimated is compared to a threshold to distinguish \ac{LoS} from \ac{NLoS} links. Second, based on the \ac{LoS} or \ac{NLoS} operation, the position of the user and blocker (if available) are calculated from the estimated parameters, setup geometry, and the refractive index of human skin. Then, based on the obtained positions and orientations, a normalized spatial blockage map for every set of snapshots is evaluated. This map's goal is to provide \acp{RIS} with a near real-time estimate of the environment dynamics.\\
\indent In our subsequent simulations, we obtain a realization of blockage score mapping while varying the system bandwidth to understand the system tradeoffs. We also compare the spectral efficiency and reliability of our system to frameworks without joint sensing and communications.
\vspace{-0.4cm}
\section{Simulation Results and Analysis}
\begin{figure}[t]
	\centering
	\includegraphics[scale=0.165]{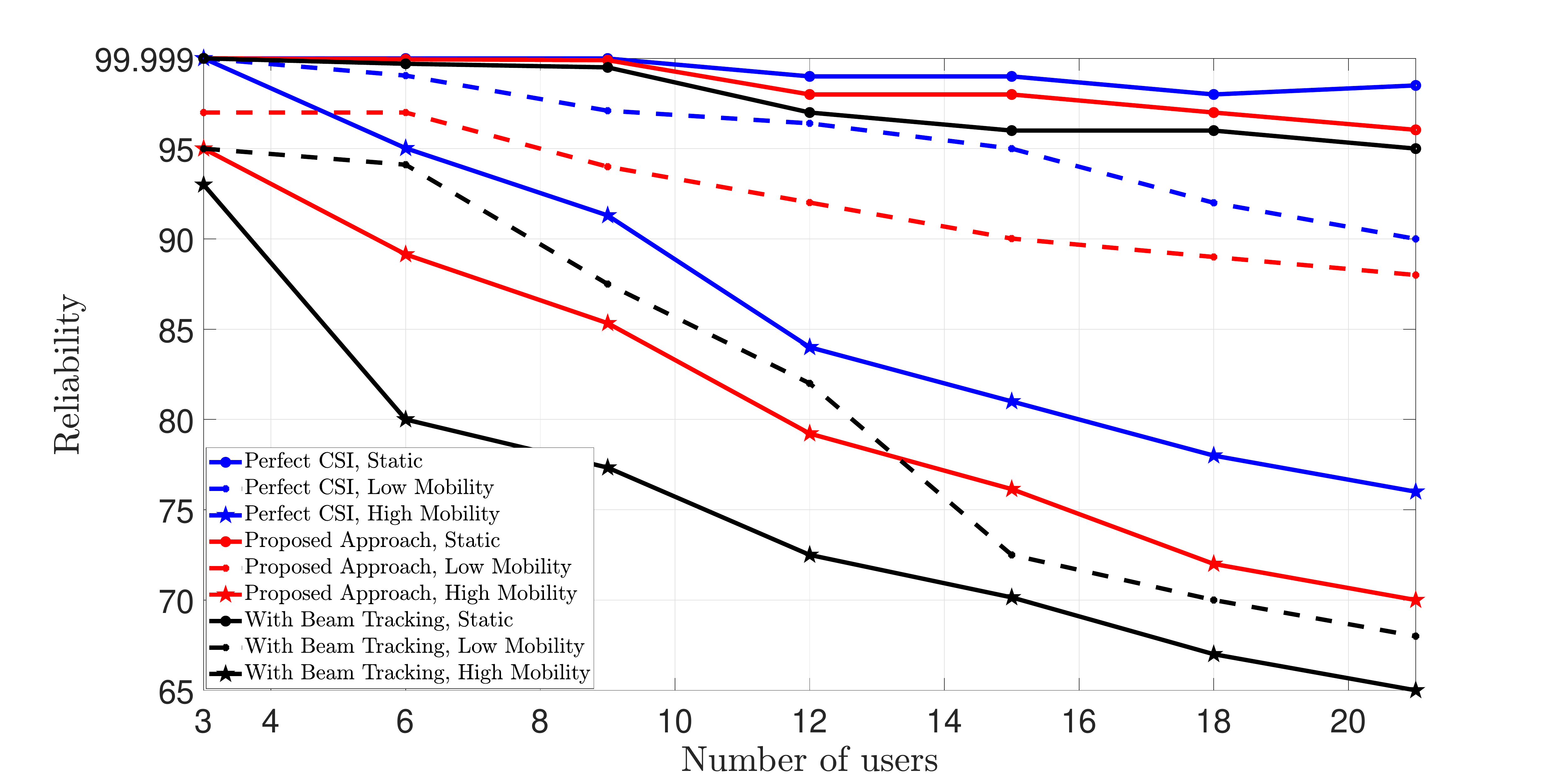}
	\vspace{-.20cm}
	\caption{\small{System reliability versus number of users}}
	\label{fig:Rel}  
	\vspace{-.4cm}
\end{figure}

\begin{figure}[t]
	\vspace{-.1cm}
	\centering
	\includegraphics[scale=0.165]{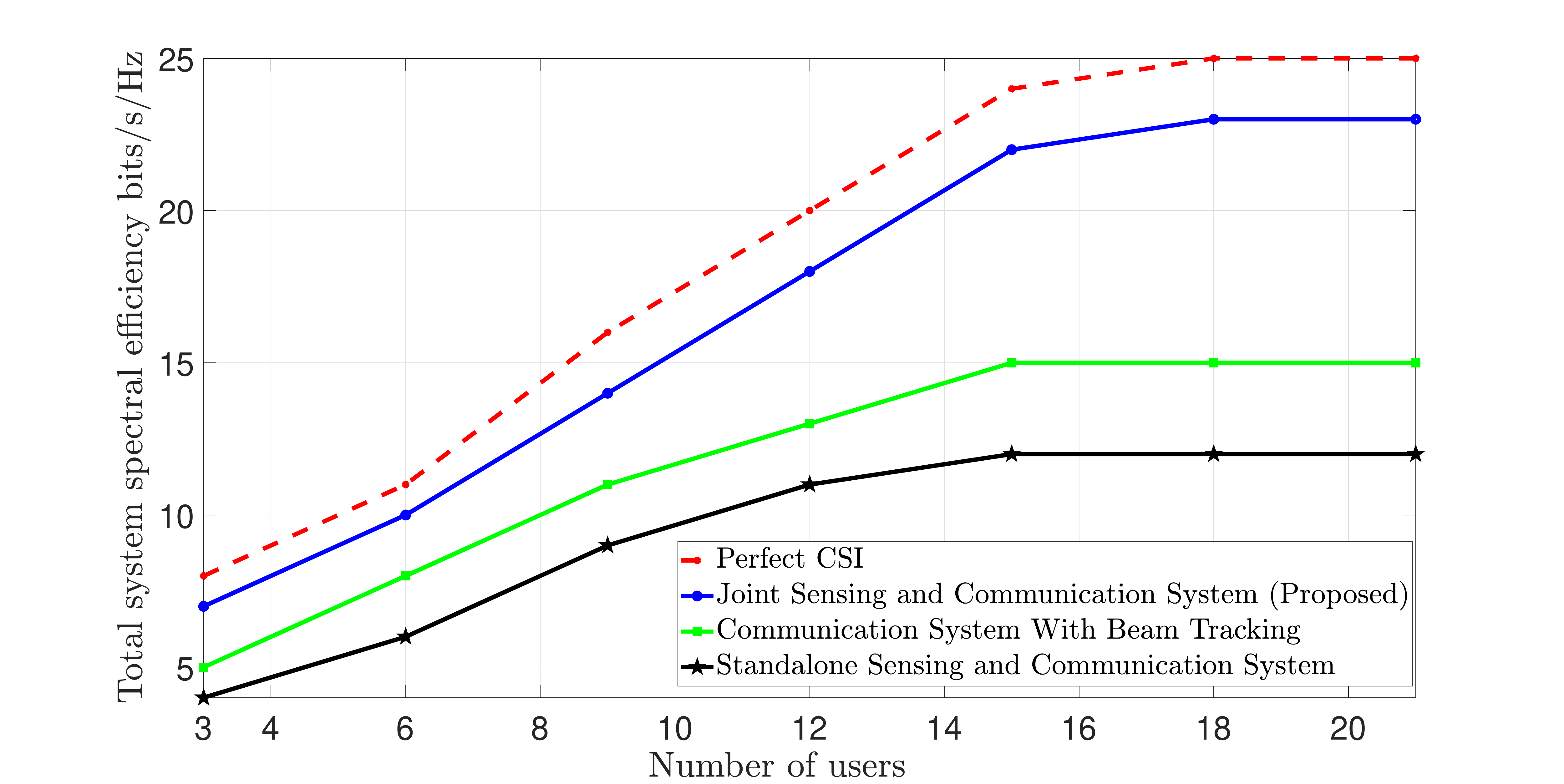}
	\vspace{-.20cm}
	\caption{\small{Spectral efficiency versus number of users}}
	\label{fig:SpectEff} 
	\vspace{-0.7cm} 
\end{figure}

For our simulations, the \acp{RIS} are deployed over the three walls of  an indoor area modeled as a square of size $\SI{24}{m}\times\SI{24}{m}$. The molecular absorption was obtained from the sub-\ac{THz} model in \cite{kokkoniemi2021line} with $1\%$ of water vapor molecules. For our parameters, we have: $N=\SI{64}{}$ antennas, $Q=\SI{32}{}$ antennas, $f=\SI{0.275}{THz}$, $W=\SI{10}{GHz}$ (unless mentioned otherwise), and  $p=\SI{30}{dBm}$. All statistical results are averaged over a large number of independent runs. The network was simulated with data generated from \ac{XR} users moving according to a random walk which constitutes the most general scheme for user mobility.\\
\indent Fig.~\ref{SpatialBlock} shows the effect of varying the bandwidth on the spatial blockage score map. That is, in the figure's close-up we can see how a higher bandwidth visualizes a more continuous range of values for the blockage score.  This is attributed to the fact that environmental sensing parameters extracted at a higher bandwidth can be computed at higher resolutions. It should be emphasized that increasing the bandwidth on the one hand, improves the data rate of our communication system, and on the other hand enhances our sensing resolution. Thus, our proposed approach does not face a \emph{rate-resolution} tradeoff. On the contrary, these two metrics are positively correlated.\\
\indent Fig.~\ref{fig:Rel} shows the system reliability versus the number of users. Here, the system reliability follows the definition in \cite{chaccour2020can}, i.e., the probability that the \ac{E2E} delay falls below a particular threshold. The \ac{E2E} delay includes both downlink an uplink delays which include the processing, queuing, and transmission time (and beam tracking when valid). The threshold used for the total \ac{E2E} delay here is $\SI{20}{ms}$. We can see that our proposed approach has a higher reliability in static, low, and high mobility. Here, low mobility refers to a walking speed of \ac{XR} users, i.e., $\SI{4.5}{km/h}$, and high mobility refers to a higher pace of $\SI{9}{km/h}$. Fig.~\ref{fig:Rel} shows that the highest gains achieved using our proposed approach is in a low mobility setup, i.e., a $\SI{10}{\%}$ improvement for the average number of users and a $\SI{19}{\%}$ improvement for $U=20$, whereby the mapping configured captures the real-time state of the environment. Moreover, in a static setup, sensing information is not very useful. In contrast, for high mobility, the sensing measurements may lag the instantaneous dynamics and thus induce delays. \\
\indent In Fig.~\ref{fig:SpectEff}, the total spectral efficiency defined as in \cite{chiriyath2017radar} is shown. Fig.~\ref{fig:SpectEff} clearly shows the benefits reaped from deploying a joint sensing and communication system that shares hardware, waveform, and spectrum. In fact, our proposed joint sensing and communication approach achieves a $\SI{42}{\%}$ and $\SI{75}{\%}$ improvement respectively compared to a communication system with beam training and a standalone sensing and communication system.
\vspace{-0.3cm}
\section{Conclusion}\label{Sec:Conclusion}
In this paper, we have proposed a new joint sensing and communication system framework for wireless \ac{XR} at \ac{THz} frequency bands. In particular, we have opportunistically used uplink communication waveforms for interleaved user and environment tracking. In particular, we  have exploited the sparsity of \ac{THz} channels to extract unique sensing parameters from the received uplink signals. Subsequently, a high-resolution indoor mapping is derived so as to characterize the spatial availability of communications and the blockage score of the area. Such a mapping allows us to reduce the need for a continuous beam tracking and improves the overall spectrum efficiency of the system.\vspace{.1cm}
\appendix
\vspace{-0.2cm}
\subsection{Proof of Proposition 1}
Problem \eqref{optimization} can be solved efficiently by the alternating least squares matrix factorization procedure \cite{sidiropoulos2017tensor}. Since we deal with a third order tensor, this factorization has three iterative steps. Then, it fixes two factor matrices and minimizes the error with respect to the considered factor matrix as:
\begin{equation*}
\small
\widetilde{\boldsymbol{A}}^{(t+1)}=\arg \min _{\widetilde{\boldsymbol{A}}}\left\|\boldsymbol{Y}_{n,(1)}^{T}-\left(\widetilde{\boldsymbol{C}}^{(t)} \odot \widetilde{\boldsymbol{B}}^{(t)}\right) \widetilde{\boldsymbol{A}}^{T}\right\|_{F}^{2}, 
\end{equation*}
\begin{equation*}
\small
\widetilde{\boldsymbol{B}}^{(t+1)}=\arg \min _{\widetilde{\boldsymbol{B}}}\left\|\boldsymbol{Y}_{n,(2)}^{T}-\left(\widetilde{\boldsymbol{C}}^{(t)} \odot \widetilde{\boldsymbol{A}}^{(t+1)}\right) \widetilde{\boldsymbol{B}}^{T}\right\|_{F}^{2}, 
\end{equation*}
\begin{equation}
\small
\widetilde{\boldsymbol{C}}^{(t+1)}=\arg \min _{\widetilde{\boldsymbol{C}}}\left\|\boldsymbol{Y}_{n,(3)}^{T}-\left(\widetilde{\boldsymbol{B}}^{(t+1)} \odot \widetilde{\boldsymbol{A}}^{(t+1)}\right) \widetilde{\boldsymbol{C}}^{T}\right\|_{F}^{2},
\end{equation}
where, $\odot$ is the Khatri-Rao product symbol. The exact factor matrices are related to their estimates according to:
$\widetilde{\boldsymbol{A}} =\boldsymbol{A} \boldsymbol{\Upsilon}_{1} \boldsymbol{\Gamma}+\boldsymbol{E}_{1}, \hspace{0.1cm}
\widetilde{\boldsymbol{B}} =\boldsymbol{B} \boldsymbol{\Upsilon}_{2} \boldsymbol{\Gamma}+\boldsymbol{E}_{2},  \hspace{0.1cm}
\widetilde{\boldsymbol{C}} =\boldsymbol{C} \boldsymbol{\Upsilon}_{3} \boldsymbol{\Gamma}+\boldsymbol{E}_{3}, $
where $\{\boldsymbol{\Upsilon_1}, \boldsymbol{\Upsilon_2}, \boldsymbol{\Upsilon_3}\}$ are unknown nonsingular diagonal matrices that satisfy $\boldsymbol{\Upsilon_1}\boldsymbol{\Upsilon_2}\boldsymbol{\Upsilon_3}=
\boldsymbol{I}$, $\boldsymbol{\Gamma}$ is an unknown permutation matrix, and $\{\boldsymbol{E_1},\boldsymbol{E_2},\boldsymbol{E_3}\}$  are estimation error matrices. By applying a maximum likelihood estimator on each of the equations and assuming that the estimation error matrices $\{\boldsymbol{E_1},\boldsymbol{E_2},\boldsymbol{E_3}\}$ follow an \ac{i.i.d.} circularly symmetric Gaussian distribution \cite{zhou2017low}, we obtain \eqref{prop1}, \eqref{prop2}, and \eqref{prop3}.
\bibliographystyle{IEEEtran}
\def\baselinestretch{0.66}
\bibliography{bibliography}
\end{document}